\documentclass[letter]{jpsj3}

\usepackage{bm}

\title{
General Properties of Response Functions of Nonequilibrium Steady States
}

\author{
Akira \textsc{Shimizu}\thanks{E-mail address: shmz@ASone.c.u-tokyo.ac.jp}
and
Tatsuro \textsc{Yuge}$^{1}$\thanks{E-mail address: yuge@m.tohoku.ac.jp} }

\inst{
Department of Basic Science, 
University of Tokyo, 
Komaba, Meguro-City, Tokyo 153-8902\\
$^{1}$IIAIR, Tohoku University, 
Aoba-ku, Sendai, Miyagi 980-8578
}

\abst{
We derive general properties, which hold for both quantum and classical 
systems, of response functions of nonequilibrium steady states.  We 
clarify differences from those of equilibrium states.  In particular, 
sum rules and asymptotic behaviors are derived, and their implications 
are discussed.  Since almost no assumptions are made, our results are 
applicable to diverse physical systems.  We also demonstrate our results 
by a molecular dynamics simulation of a many-body interacting system.
}

\kword{sum rule, asymptotic behavior, reciprocal relation, fluctuation-dissipation relation}

\begin{document}

\maketitle

Stable states of macroscopic systems can be well characterized 
by their responses to external probe fields.
When the probe fields are weak, the responses are linear functions of 
the probe fields.
General properties of the linear response functions 
are well-known 
for equilibrium states \cite{KTH,Zubarev}.
In contrast, those for nonequilibrium steady states (NESSs) 
are not well understood yet,
although many attempts have been made
\cite{keldysh,Shen,ON1988,BMW2009,AG2007,SU92,YS2009,HaradaSasa,YIS,YS2007,YS_alder}.

For example, 
perturbation expansion of the density operator $\hat{\rho}_F$ of NESSs
in terms of a driving field $F$ 
has often been employed \cite{Shen}.
This gives linear and higher-order ($n=2, 3, \cdots$)
response functions (denoted by $\Phi_{\rm eq}$ and 
$\Phi^{(n)}_{\rm eq}$, respectively) of {\em equilibrium states}.
General properties, such as symmetries, of $\Phi^{(n)}_{\rm eq}$
were thus derived \cite{Shen}.
Similar results were also obtained from 
the fluctuation theorems\cite{AG2007}.
However, it is generally hard to obtain 
linear response functions $\Phi_F$ of NESSs from 
$\Phi_{\rm eq}$ and $\Phi^{(n)}_{\rm eq}$
because such expansion converges only slowly 
(or does not converge)
for $F$ that is large enough to drive NESSs of interest.
Another approach is to utilize some general expression of $\Phi_F$.
Such an expression 
was derived, 
e.g., in ref.~\citen{BMW2009}.
However, 
it contains the expectation value of a function 
which is unknown 
except for simple cases \cite{note:C}.
To derive physical results for $\Phi_F$ from
such a formal expression,
simplifying assumptions were made\cite{BMW2009}, 
at the expense of generality.
Furthermore, 
one may expect that 
$\Phi_F$ 
could be expressed by small fluctuation in NESSs.
However, refs.~\citen{SU92} and \citen{YS2009} showed that 
$\Phi_F$ of finite macroscopic systems is {\em not} a universal function of 
the fluctuation and temperature, i.e., $\Phi_F$ depends 
also on another system-dependent parameter(s).
Because of these difficulties, 
general properties of $\Phi_F$ were not clarified.

In this paper, 
we derive general properties of 
$\Phi_F$, 
which hold for diverse physical systems.
We clarify which properties 
are common or different between 
$\Phi_F$ and $\Phi_{\rm eq}$.
We also illustrate some of the properties 
by a molecular dynamics (MD) simulation of a
many-body 
system.

{\em Response function of NESS -- }
Suppose that a strong static field $F$ is applied to the target system 
(the macroscopic system of interest), 
and a NESS 
is realized
for a sufficiently long time, i.e., for $[t_{\rm in}, t_{\rm out}]$,
where $t_{\rm out} - t_{\rm in}$ is macroscopically long. 
In such a NESS 
every macroscopic variable $A$ takes a constant 
value $\langle A \rangle_F$ 
in the sense that 
its expectation value at time $t$ behaves as
\begin{equation}
\langle A \rangle_F^t = \langle A \rangle_F + 
o\left( \langle A \rangle_{\rm tp} \right).
\label{eq:steady}\end{equation}
Here, 
$\langle A \rangle_{\rm tp}$ denotes a typical value of $A$,
and 
$o\left( \langle A \rangle_{\rm tp} \right)$ represents
a (time-dependent) term which is negligibly small in the sense that 
$o\left( \langle A \rangle_{\rm tp} \right) / \langle A \rangle_{\rm tp}
\to 0$ as $V \to \infty$,
where $V$ denotes the volume of the target system.
When $A$ is the energy $U$, for example, 
$\langle U \rangle_{\rm tp} = O(V)$ and
$\langle U \rangle_F^t = \langle U \rangle_F + o\left( V \right)$.

Suppose that a weak and time-dependent {\em probe field} $f(t)$ is applied, 
in addition to $F$, 
to the target system for $t \geq t_0$, 
where $t_{\rm in} < t_0 \leq t < t_{\rm out}$.
We are interested in the response of the NESS to $f(t)$.
Specifically, we focus on the response,
\begin{equation}
\Delta A(t)
\equiv 
\langle A \rangle_{F+f}^t - \langle A \rangle_F,
\end{equation}
of a macroscopic variable $A$ of the target system.
To the linear order in $f$, $\Delta A(t)$ can be expressed as
\begin{equation}
\Delta A(t)
=\int_{t_0}^t \Phi_F(t-t') f(t') dt',
\label{eq:linear_response}\end{equation}
which we call the {\em linear response relation}.
This and the causality relation
\begin{equation}
\Phi_F(\tau) = 0 \mbox{ for } \tau<0,
\label{eq:causality}\end{equation}
define the {\em response function} $\Phi_F(\tau)$ of the NESS,
as in the case of $\Phi_{\rm eq}(\tau)$ \cite{KTH,Zubarev}.

{\em Microscopic expression of $\Phi_F$ -- }
Equations (\ref{eq:linear_response}) and (\ref{eq:causality}) do not 
refer to microscopic physics at all -- they are phenomenological 
equations which are closed in a macroscopic level. 
We now relate them to microscopic physics by 
deriving a microscopic expression of $\Phi_F(\tau)$.

\begin{figure}
\begin{center}
\includegraphics[width=.6\linewidth]{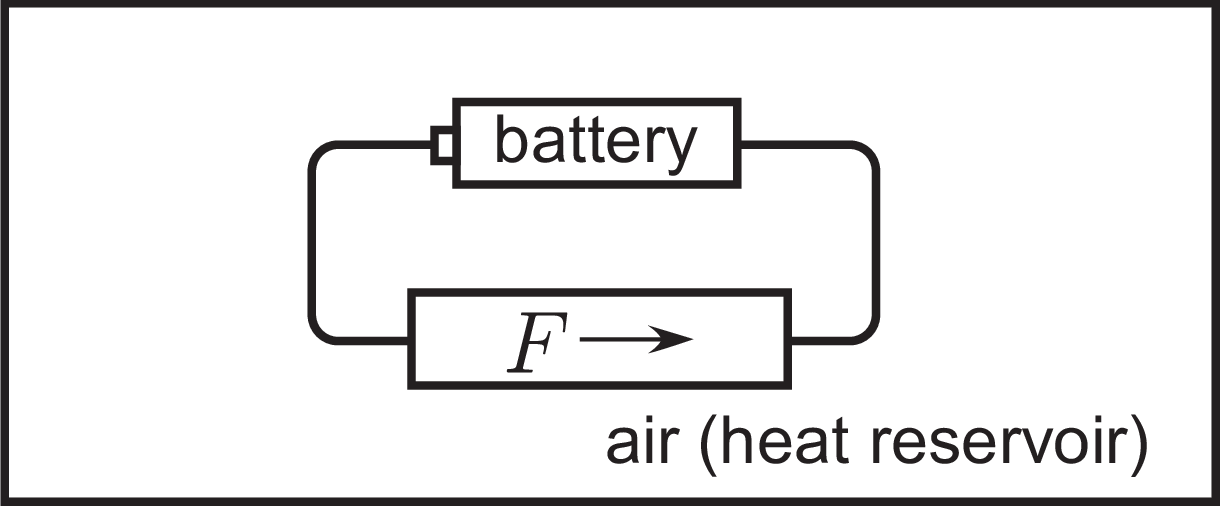}
\end{center}
\vspace{-3mm}
\caption{
An example of a large system which we call the total system.
It includes an electrical conductor, a battery, a heat reservoir, 
and so on.
}
\label{fig:closed_NESS}
\end{figure}

Since we are interested in {\em general} properties of 
NESSs, 
we do {\em not} employ perturbation expansion with respect to $F$
\cite{Shen,AG2007},
which, for large $|F|$ of interest, 
converges only slowly or does not converge
except 
in limited physical situations.
To treat $F$ non-perturbatively,
we consider a large system which includes 
the target system, a driving source that generates $F$, 
and a heat reservoir(s).
We call this large system the {\em total system}, and 
denote its Hamiltonian by $\hat{H}^{\rm tot}$.
When the target system is an electrical conductor, for example, 
the driving source may be a battery, 
the heat reservoir may be the air,  
and the total system is the one that includes them all,
as shown in Fig.~\ref{fig:closed_NESS}.
On the other hand, we do {\em not} include 
the source of the probe field $f$, such as a microwave generator,
in the total system. 
We assume that $f$ gives rise to the interaction term
$- \hat{B} f(t)$, where $\hat{B}$ is a macroscopic variable of the 
target system.
Hence, the total system is an isolated system except that 
it is subject to an external weak field $f$.
Therefore, the density operator
of the total system $\hat{\rho}^{\rm tot}_{F+f}(t)$ evolves as
\begin{equation}
i \hbar {\partial \over \partial t} \hat{\rho}^{\rm tot}_{F+f}(t)
= \left[ \hat{H}^{\rm tot} - \hat{B} f(t),\ 
\hat{\rho}^{\rm tot}_{F+f}(t) \right].
\label{eq:vN_rho_tot}\end{equation}

We denote $\hat{\rho}^{\rm tot}_{F+f}(t)$
with $f=0$ by $\hat{\rho}^{\rm tot}_{F}(t)$.
When $f=0$, 
the reduced density operator of the target system is 
\begin{equation}
\hat{\rho}_F
\equiv
{\rm Tr}' \left[ \hat{\rho}^{\rm tot}_{F}(t) \right],
\end{equation}
where ${\rm Tr}'$ denotes the trace operation over 
the degrees of freedom other than those of the target system.
For a time interval $[t_{\rm in}, t_{\rm out}]$,
a NESS is realized in the target system
(while the driving source such as a battery is not in a steady state),
and, according to eq.~(\ref{eq:steady}),
we can regard $\hat{\rho}_F$ as being independent of $t$
as far as macroscopic variables are concerned.
Unlike the equilibrium case, however, an explicit form of 
$\hat \rho_F$ is unknown.

When $f \neq 0$, 
$\hat{\rho}^{\rm tot}_{F}(t)$ 
is changed into 
$\hat{\rho}^{\rm tot}_{F+f}(t)$.
We {\em assume that the NESS is stable against small perturbations}.
That is, after small perturbations are removed
the target system returns to the same NESS 
as that before they were applied.
Except for NESSs of some soft matters and glass near a melting point,
most NESSs,
including those of 
nonlinear optical materials \cite{Shen,ON1988}
and electrical conductors, 
satisfy this assumption.

For such NESSs, we can evaluate $\Phi_F$ 
by evaluating the solution of eq.~(\ref{eq:vN_rho_tot})
using a first-order perturbation expansion with respect to $f$.
For an observable of interest, $\hat{A}$, of the target system,
its response
$
\Delta A(t)
\equiv
\mbox{Tr} [ \hat{\rho}^{\rm tot}_{F+f}(t) \hat{A} ] 
- 
\mbox{Tr} [ \hat{\rho}^{\rm tot}_{F}(t_0) \hat{A} ]
=
\mbox{Tr} [ \hat{\rho}^{\rm tot}_{F+f}(t) \hat{A} ] 
- 
\mbox{Tr} [ \hat \rho_F \hat{A} ]
$
is evaluated to the linear order in $f$ as
\begin{equation}
\Delta A(t)
=
\int_{t_0}^t 
{1 \over i \hbar}
\mbox{Tr}
\left(
\hat{\rho}^{\rm tot}_{F}(t)
\left[\breve{B}(t'-t), \
\hat{A}
\right]
%
%
\right)
f(t') dt',
\label{eq:DeltaA_soboku}\end{equation}
where the symbol ` $\breve\null$ ' denotes an operator 
in the interaction picture, i.e., 
\[
\breve B(t) \equiv 
\exp \left({i \over \hbar} \hat{H}^{\rm tot} t \right) \hat{B} \, 
\exp \left({-i \over \hbar} \hat{H}^{\rm tot} t \right).
%
%
\]
From consistency with the macroscopic physics, eq.~(\ref{eq:linear_response}), 
$t$ in $\hat{\rho}^{\rm tot}_{F}(t)$ in eq.~(\ref{eq:DeltaA_soboku})
must be irrelevant.
Hence, we can take $t$ to be an arbitrary time (such as $t_0$)
in $[t_{\rm in}, t_{\rm out}]$,
and simply write $\hat{\rho}^{\rm tot}_{F}(t)$ as $\hat{\rho}^{\rm tot}_{F}$.
We thus obtain a general formula; 
\begin{equation}
\Phi^{AB}_F(\tau)
=
{1 \over i \hbar}
\mbox{Tr}
\left(
\hat{\rho}^{\rm tot}_{F}
\left[ \breve{B}(-\tau), \ \hat{A} \right]
%
%
\right)
\mbox{ for } \tau \geq 0.
\label{eq:RCR}\end{equation}
Here, 
we denote $\Phi_F(\tau)$ by $\Phi^{AB}_F(\tau)$ to designate
variables $A$ and $B$.
For classical systems,
$
[ \breve{B}(-\tau), \ \hat{A} ]/i \hbar
$
(and similar expressions in the following equations)
should be replaced with the corresponding Poisson bracket.

The right-hand side (rhs) of eq.~(\ref{eq:RCR}) 
represents some correlation in the NESS.
If it could reduce to the symmetrized time correlation\cite{KTH}, 
it would be equivalent to fluctuation 
(in the classical regime, $k_B T \gg \hbar \omega$\cite{note:B}).
However, as will be discussed shortly, this is not the case when $F \neq 0$.
Hence, 
we do {\em not} 
call eq.~(\ref{eq:RCR}) a fluctuation-dissipation relation (FDR).
We call it 
the {\em response-correlation relation} (RCR).
Among similar formulas, 
eq.~(\ref{eq:RCR}) has the most convenient form 
to derive useful properties which will be presented below. 

{\em Fluctuation-dissipation and reciprocal relations -- }
Before deriving useful properties, 
we use the RCR 
to discuss why the FDR 
and 
the reciprocal relations
(including those for 
finite frequencies \cite{KTH}) 
are violated in NESSs\cite{SU92,YS2009}.

When $F =0$, 
$\hat{\rho}^{\rm tot}_{F}$ and $\Phi^{AB}_F$
of eq.~(\ref{eq:RCR}) reduce to 
the equilibrium state $\hat \rho_{\rm eq}$ 
and its response function $\Phi^{AB}_{\rm eq}$, respectively,
and the RCR reduces to the equilibrium one\cite{KTH,Zubarev}
(which is not customarily called the RCR, though). 
When the canonical ensemble, e.g., is employed, 
both $\hat \rho_{\rm eq} \propto e^{-\beta \hat{H}^{\rm tot}}$ 
and $e^{{i \over \hbar} \hat{H}^{\rm tot} t}$
(which defines $\breve A(\tau)$) are exponential functions of 
$\hat{H}^{\rm tot}$.
Using this fact, 
we can recast the equilibrium RCR as
\begin{equation}
\Phi_{\rm eq}^{AB}(\tau)
={1 \over k_B T} \langle \dot{\breve{B}}(0); \breve  A(\tau) \rangle _{\rm eq},
\label{eq:Kubo_formula}\end{equation}
where $\langle  \, \cdot \, ; \, \cdot \, \rangle _{\rm eq}$ 
denotes the canonical correlation \cite{KTH}.
This result is known as the Kubo formula, from which 
one can derive the reciprocal relations \cite{KTH,Zubarev}.
In the classical regime ($k_B T \gg \hbar \omega$), 
the canonical correlation reduces to the symmetrized 
time correlation\cite{KTH},
and hence to fluctuation\cite{note:B}, 
and one obtains the FDR \cite{KTH,Zubarev}.
%
%

When $F \neq 0$, in contrast, 
$\hat{\rho}^{\rm tot}_F$ (although its explicit form is unknown)
cannot be an exponential function of 
$\hat{H}^{\rm tot}$ only.
As a result, 
the RCR cannot be rewritten into a form similar to 
eq.~(\ref{eq:Kubo_formula}).
That is, 
the RCR holds both for equilibrium states and for NESSs, 
whereas it is equivalent to the Kubo formula only 
for the former. 
As a result, the FDR and the reciprocal relations 
are violated in NESSs.
The difference between the rhs of eq.~(\ref{eq:RCR}) 
and the symmetrized time correlation divided by $k_B T$
is the violating term.

{\em Properties derived from the phenomenological equations -- }
Equations (\ref{eq:linear_response}) and (\ref{eq:causality})
take the same forms as those for $\Phi_{\rm eq}$. 
Therefore, among many properties of $\Phi_{\rm eq}$, those derivable 
only from eqs.~(\ref{eq:linear_response}) and (\ref{eq:causality})
(without using the Kubo formula) 
hold also for $\Phi_F$.
For completeness, 
we mention such properties, although some of them may be rather obvious.

For stable NESSs, we expect that $\left| \Phi_F(\tau) \right|$ is integrable 
over $(-\infty, +\infty)$ \cite{range_of_tau}.
Hence, the Fourier transform
\begin{equation}
\Xi_F(\omega)
\equiv
\int_{-\infty}^\infty \Phi_F(\tau) e^{i \omega \tau} d\tau
=
\int_{0}^\infty \Phi_F(\tau) e^{i \omega \tau} d\tau
\label{def:Xi}\end{equation}
should be a continuous function of $\omega$.
As in the case of the Fourier transform $\Xi_{\rm eq}(\omega)$ of
$\Phi_{\rm eq}(\tau)$, 
we can easily show that 
$\Xi_F(\omega)$ satisfies the dispersion relations,
\begin{eqnarray}
{\rm Re} \, \Xi_F(\omega) 
&=& 
\int_{-\infty}^{\infty} {{\cal P} \over \omega'- \omega} 
{\rm Im} \, \Xi_F(\omega') {d \omega' \over \pi},
\label{eq:DR1}\\
{\rm Im} \, \Xi_F(\omega) 
&=& 
- \int_{-\infty}^{\infty} {{\cal P} \over \omega'- \omega} 
{\rm Re} \, \Xi_F(\omega') {d \omega' \over \pi},
\label{eq:DR2}\end{eqnarray}
and the moment sum rules \cite{KTH}.
We also see that 
${\rm Re} \, \Xi_F(\omega)$ is even, whereas ${\rm Im} \, \Xi_F(\omega)$ 
is odd.

{\em Properties derived from the RCR -- }
We now present the most important results of this paper.
For $F=0$, 
many properties 
were previously derived for $\Phi_{\rm eq}$ 
from the Kubo formula \cite{KTH,Zubarev}.
As discussed above, 
some of them  (such as the FDR)
are violated for $\Phi_F$ when $F \neq 0$.
However, the other properties of $\Phi_{\rm eq}$ can 
actually be derived from the RCR
without using the Kubo formula, 
although they were often derived from the Kubo formula
(or similar expressions) in the literature.
Such properties hold also for $\Phi_F$ 
if some quantities are replaced with those of a NESS (see below)
because
the RCR holds even when $F \neq 0$.
We now present them.

Note that although their {\em forms} are similar to 
those for $\Phi_{\rm eq}$ \cite{KTH,Zubarev}, 
their {\em values} are often different 
from those of $\Phi_{\rm eq}$, 
as will be illustrated later.

Integration of $\Xi^{AB}_F(\omega)$ yields
\begin{equation}
\int_{-\infty}^{\infty} \Xi^{AB}_F(\omega) {d \omega \over \pi}
=
\Phi^{AB}_F(+0)
=
{1 \over i \hbar} {\rm Tr} \left( 
\hat{\rho}^{\rm tot}_{F} \left[ \breve{B}(0),  \hat{A} \right] 
\right).
\nonumber\end{equation}
Since 
$\breve B(0)$ ($=\hat{B}$) and $\hat{A}$ 
are operators of 
the target system, 
\begin{equation}
{\rm Tr} \left( 
\hat{\rho}^{\rm tot}_{F} \left[ \breve{B}(0),  \hat{A} \right] 
\right)
=
{\rm Tr} \left(
\hat{\rho}_F \left[ \hat{B}, \hat{A} \right]
\right)
=
\left\langle \left[ \hat{B}, \hat{A} \right] \right\rangle _F,
\nonumber\end{equation}
where
$
\langle \cdot \rangle _F 
\equiv {\rm Tr} \left( \hat \rho_F \ \cdot \ \right)
$
denotes the expectation value in the NESS.
Noting also that ${\rm Im} \, \Xi^{AB}_F(\omega)$ is an odd function, 
we obtain 
the following 
sum rule for ${\rm Re} \, \Xi^{AB}_F$;
\begin{equation}
\int_{-\infty}^{\infty} {\rm Re} \, \Xi^{AB}_F(\omega) {d \omega \over \pi}
=
\left\langle  
{1 \over i \hbar} \left[ \hat{B}, \hat{A} \right]
\right\rangle _F.
\label{sr:ReXi}\end{equation}
Since the rhs is the expectation value
of a {\em known} operator, it can  
{\em easily be measured experimentally} \cite{note:C}.

Moreover, by integrating eq.~(\ref{def:Xi}) by parts, 
multiplying the result with $\omega$,
and integrating the resultant equation, 
we obtain the following sum rule for ${\rm Im} \, \Xi^{AB}_F$;
\begin{eqnarray}
&& 
\int_{-\infty}^{\infty} 
\left\{ 
\omega \, {\rm Im} \, \Xi^{AB}_F(\omega) 
-
\left\langle  
{1 \over i \hbar} \left[ \hat{B}, \hat{A} \right] 
\right\rangle _F 
\right\}
{d \omega \over \pi}
\nonumber\\
&& \quad
=
- \left\langle  
{1 \over i \hbar} \left[ \dot{\breve{B}}(0), \hat{A} \right]
\right\rangle _F.
\label{sr:ImXi}\end{eqnarray}
In the second line, we have replaced 
$\hat{\rho}^{\rm tot}_F$ with $\hat{\rho}_F$ because
$\dot{\breve{B}}(0)$ 
($=[ \hat{B}, \hat{H}^{\rm tot} ]/i \hbar$)
%
%
is localized in the target system.
(Recall that all physical interactions in $\hat{H}^{\rm tot}$
should be local interactions.)
The expectation values in this sum rule
can also be measured experimentally.

From these sum rules 
we can see the asymptotic behavior of $\Xi^{AB}_F(\omega)$.
As $|\omega|$ is increased, $\Xi^{AB}_F(\omega)$ 
should decay quickly enough such that the integrals of 
eqs.~(\ref{sr:ReXi}) and (\ref{sr:ImXi}) converge.
In particular, we find 
\begin{equation}
\lim_{\omega \to \infty} \omega \, {\rm Im} \, \Xi^{AB}_F(\omega) 
=
\left\langle  
{1 \over i \hbar} \left[ \hat{B}, \hat{A} \right]
\right\rangle _F.
\label{asymptotic}\end{equation}

As discussed above,
the reciprocal relation for $\Phi^{AB}_{\rm eq}(t)$
does not hold for $\Phi^{AB}_F(t)$, 
i.e., for $\Xi^{AB}_F(\omega)$ for {\em each} $\omega$. 
However, eq.~(\ref{sr:ReXi}) yields
\begin{equation}
\int_{-\infty}^{\infty} {\rm Re} \, \Xi^{AB}_F(\omega) d \omega
=
- \int_{-\infty}^{\infty} {\rm Re} \, \Xi^{BA}_F(\omega) d \omega,
\label{reciprocal}\end{equation}
i.e., 
a reciprocal relation holds for the {\em integrated} values.

{\em Implications -- }
The left-hand side of eq.~(\ref{sr:ReXi})
equals to $\Phi^{AB}_F(+0)$ (representing instantaneous response),
which however is hardly measurable in real physical systems.
In contrast, 
${\rm Re} \, \Xi^{AB}_F(\omega)$ is 
measurable in
a certain finite range of $\omega$.
For higher $\omega$, which is out of such a range, 
${\rm Re} \, \Xi^{AB}_F(\omega)$ decays quickly, as mentioned above.
Hence, one does not necessarily have to measure it for higher $\omega$.
Therefore, eq.~(\ref{sr:ReXi}) 
should be considered as a prediction on 
${\rm Re} \, \Xi^{AB}_F(\omega)$ in a certain finite range of $\omega$.
For similar reasons, 
eqs.~(\ref{sr:ImXi})-(\ref{reciprocal})
become important 
when one wants to get information on $\Xi^{AB}_F(\omega)$.

These equations, like the corresponding ones for $\Xi^{AB}_{\rm eq}(\omega)$, 
are very useful for 
measuring or theoretically calculating $\Xi^{AB}_F(\omega)$.
For example, 
one can check experimental or theoretical results against them.
Using them and eqs.~(\ref{eq:DR1}) and (\ref{eq:DR2}), 
one can also estimate $\Xi^{AB}_F(\omega)$ 
in some range of $\omega$ 
from existing data of $\Xi^{AB}_F(\omega)$
in another range.
Moreover, as will be illustrated for a Langevin model later, 
our results can show that some equality is identical to 
another equality, which were previously treated as independent 
equalities.

Furthermore, we can see the following.
According to eq.~(\ref{sr:ReXi}), the sum value (integral)
of ${\rm Re} \, \Xi_F(\omega)$ equals to 
the expectation value $\langle C \rangle _F$ of the Hermitian operator
$ 
\hat{C} \equiv 
[ \hat{B}, \hat{A} ]/i \hbar.
$ 
Since this is an equal-time commutator, 
$\hat{C}$ depends neither on the Hamiltonian nor on the state.
There is no difference in $\hat{C}$ between 
free particles and interacting ones 
or between equilibrium states and NESSs.
Only through $\hat{\rho}_F$ the sum value
can be affected by these factors.

When $\hat{A}$ and $\hat{B}$ are linear functions of canonical variables, 
in particular, 
$\hat{C} \propto \hat{1}$ (identity operator)
and hence the sum takes the same value for {\em every} state.
More generally, we can say the same when 
$\mbox{Tr} [ \hat{\rho}^{\rm tot}_{F}(t) \hat{C} ]$ 
is conserved during evolution from 
an equilibrium sate to NESSs of interest.

For example, suppose that the target system is an electrical conductor 
of length $L$.
A static electric field $F$ is applied 
in the $x$ direction (along the conductor).
Let $\hat{q}_x^j$ and $\hat{p}_x^j$ be the $x$ components of the position and
momentum, respectively, of the $j$th electron in the conductor.
Then, $\hat{B} = \sum_{j} e \, \hat{q}_x^j$,
and the electric current averaged over the $x$ direction may be given by
\begin{equation}
\hat{I} \equiv {1 \over L} \sum_{j} {e \over m} \, \hat{p}_x^j,
\label{current}\end{equation}
where $m$ is electron's mass.
If one is interested in $\hat{I}$, putting
$\hat{A} = \hat{I}$ yields $\hat{C} = (e^2 N_e/mL) \hat{1}$,
where $N_e$ is the number of electrons in the conductor.
We thus find 
\begin{equation}
\int_{-\infty}^{\infty} {\rm Re} \, \Xi^{I B}_F(\omega) {d \omega \over \pi}
=
{e^2 N_e \over m L},
\label{sumReXiIB}\end{equation}
which is independent of $F$.
Generally, ${\rm Re} \, \Xi^{I B}_F(\omega)$
at low $\omega$ depends strongly on $F$ for large $|F|$.
At high $\omega$, on the other hand, it is expected that 
${\rm Re} \, \Xi^{I B}_F(\omega)$ would be insensitive to $F$
because each particle would not collide with other particles
in a short time period $\sim 1/\omega$.
(These facts will be illustrated later.)
From these viewpoints, 
eq.~(\ref{sumReXiIB})
may be counterintuitive, and therefore is useful.

Regarding ${\rm Im} \, \Xi^{IB}_F$, we can apply eqs.~(\ref{sr:ImXi})
and (\ref{asymptotic}). For example, the latter yields
\begin{equation}
\lim_{\omega \to \infty} 
\omega \, {\rm Im} \, \Xi^{IB}_F(\omega) 
=
{e^2 N_e \over m L},
\label{asymptotic_IB}\end{equation}
which is also independent of $F$.

For more general cases where 
$\mbox{Tr} [ \hat{\rho}^{\rm tot}_{F}(t) \hat{C} ]$ 
is not conserved
during evolution from an equilibrium sate to NESSs of interest,
the sum value generally depends on $F$.
For example, 
if one is interested in $\hat{I}^2$ (to investigate, e.g., current fluctuation)
in the above example, putting
$\hat{A} = \hat{I}^2$ yields
$\hat{C} = (2 e^2 N_e/mL) \hat{I}$.
We thus find
\begin{equation}
\int_{-\infty}^{\infty} {\rm Re} \, \Xi^{I^2 B}_F(\omega) {d \omega \over \pi}
=
\lim_{\omega \to \infty}
\omega \, {\rm Im} \, \Xi^{I^2 B}_F(\omega) 
=
{2 e^2 N_e \over m L} \langle I \rangle _F,
\label{sumReXiI^2B}\end{equation}
which depends strongly on $F$.
This fact demonstrates that 
although the {\em forms} of eqs.~(\ref{sr:ReXi}) and (\ref{asymptotic})
are similar to the corresponding ones 
for $\Phi_{\rm eq}$ \cite{KTH,Zubarev}, 
their {\em values} can be very different.

{\em Non-Hamiltonian systems -- }
We have assumed 
that the total system, such as Fig.~\ref{fig:closed_NESS}, is a 
Hamiltonian system. 
In studies of NESSs, 
non-Hamiltonian models, such as stochastic models, are often employed.
The general properties of $\Xi^{A B}_F$ must hold also in such models
{\em if the models are physically reasonable ones}, 
because every existing physical system is believed to be 
a Hamiltonian system if a sufficiently large system 
(such as Fig.~\ref{fig:closed_NESS}) is considered.

For example, a nonlinear Langevin model
\begin{equation}
\dot{p} = - (\gamma/m) p - U'(x) + F + f(t) + \xi(t),
\label{eq:Langevin}\end{equation}
where $p=m \dot{x}$ and $U(x)$ is a potential, may be derived from 
a Hamiltonian model by making projection and by
approximating $\gamma$ and $\xi(t)$ 
as a constant and white noise, respectively.
If these approximations are physically reasonable, 
eq.~(\ref{sr:ReXi}) must hold in this Langevin model 
because it holds in the original
Hamiltonian model.
We can prove that this is the case
for any value of $F$.
Hence, the Langevin model is physically reasonable in view of 
eq.~(\ref{sr:ReXi}).
On the other hand, it is well-known that 
the Langevin model gets worse as $|F|$ is increased.
This illustrates that eq.~(\ref{sr:ReXi}), 
and the other general properties derived above, 
are {\em not sufficient but necessary conditions} 
for good nonequilibrium models.
In this respect, they are similar to the charge conservation,
which is also a universal and necessary condition for good models.

As an illustration of significance of eq.~(\ref{sr:ReXi}) on the 
Langevin model, we can show using eq.~(\ref{sr:ReXi}) that  
equality (5) of ref.~\citen{HaradaSasa} on dissipation 
$\langle J \rangle_F$ is identical to a simple relation,
$\langle J \rangle_F = (\gamma/m) (\langle p^2 \rangle_F/m - k_B T)$,
of refs.~\citen{HN79} [eq.~(82)] and \citen{SekimotoBOOK} [eq.~(4.13)].

{\em Numerical example -- }
Finally, we illustrate the validity of eqs.~(\ref{sumReXiIB}) 
and (\ref{asymptotic_IB})
by an MD simulation of a model of 
a classical two-dimensional electrical conductor \cite{YIS,YS2007,YS_alder,YS2009}.
The model includes $N_e$ electrons ({\it e}), $N_p$ phonons ({\it p}) 
and $N_i$ impurities ({\it i}), where $N_e, N_p, N_i \gg 1$.
The {\it e-e}, {\it e-p}, {\it e-i}, {\it p-p} and {\it p-i}
interactions are all present, whereas 
the static electric field $F$ acts only on electrons.
The energy of this many-body system is dissipated through 
thermal walls 
(which simulate a heat reservoir) 
for phonons, and a NESS is realized for each value of $F$.
This model is a mechanical model supplemented by 
the thermal walls for phonons.
For NESSs at large $|F|$, we previously found the following:
(i) $\langle I \rangle _F$ is nonlinear in $F$ \cite{YIS},
(ii) the long-time tail is strongly modified \cite{YS2007,YS_alder},
and
(iii) the FDR is significantly violated \cite{YS2009}.

To compute $\Xi^{I B}_F(\omega)$,
we take the probe field $f(t)$ to be monochromatic; 
$f(t) \propto \sin (\omega t)$.
We apply it in the $x$ direction in addition to $F$, and 
perform an MD simulation,
in which we here take $e=m=1, N_e=N_p=1500, N_i=500$, and $L=750$.
By calculating the classical counterpart of $\hat{I}(t)$ of eq.~(\ref{current})
for a sufficiently long time ($\gg 1/\omega$), 
we obtain 
$\Xi^{I B}_F(\omega)$.
This procedure is repeated for various values of 
$\omega$ and $F$.

\begin{figure}
\begin{center}
\includegraphics[width=.8\linewidth]{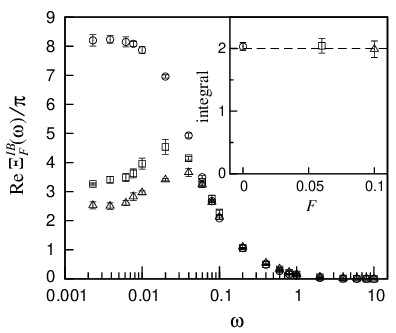}
\end{center}
\vspace{-5mm}
\caption{
${\rm Re} \, \Xi^{I B}_F(\omega) / \pi$ for $F=0$ (circles), $0.06$ (squares) 
and $0.1$ (triangles).
The inset shows the $\omega$-integral of 
${\rm Re} \, \Xi^{I B}_F(\omega) / \pi$ 
plotted against $F$, where
the dashed line represents 
the rhs of eq.~(\ref{sumReXiIB}).
}
\label{fig:ReXiIB}
\end{figure}

Figure~\ref{fig:ReXiIB} shows $\omega$-dependence
of ${\rm Re} \, \Xi^{I B}_F(\omega) / \pi$ for three different values of $F$. 
At low frequencies ${\rm Re} \, \Xi^{I B}_F(\omega)$ depends strongly on $F$,
implying that the response to $F$ is nonlinear.
At high frequencies, the $F$-dependence looks quite weak.
However, since 
the horizontal axis is in the logarithmic scale and
$\int \Xi(\omega) d \omega = \int \omega \, \Xi(\omega) d (\ln \omega)$, 
small differences in ${\rm Re} \, \Xi^{I B}_F(\omega)$ at 
high frequencies 
contribute significantly to the 
$\omega$-integral of eq.~(\ref{sumReXiIB}).
As a result, the integral over all $\omega$
(i.e., the left-hand side of eq.~(\ref{sumReXiIB}))
becomes independent of $F$ (within possible numerical errors),
as shown in the inset of Fig.~\ref{fig:ReXiIB}.
Moreover, the value of the integral agrees with 
that predicted by 
eq.~(\ref{sumReXiIB}).

\begin{figure}
\begin{center}
\includegraphics[width=.8\linewidth]{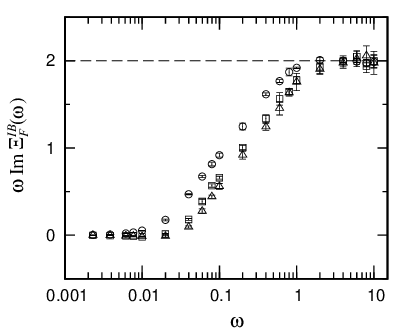}
\end{center}
\vspace{-5mm}
\caption{
$\omega~{\rm Im} \, \Xi^{I B}_F(\omega)$ for $F=0$ (circles), $0.06$ (squares) 
and $0.1$ (triangles).
The dashed line represents 
the rhs of eq.~(\ref{asymptotic_IB}).
}
\label{fig:ImXiIB}
\end{figure}

Furthermore, Fig.~\ref{fig:ImXiIB} shows $\omega$-dependence
of $\omega \, {\rm Im} \, \Xi^{I B}_F(\omega)$ for 
three different values of $F$.
We observe that as $\omega$ is increased 
$\omega \, {\rm Im} \, \Xi^{I B}_F(\omega)$ 
approaches 
the same asymptotic value,
which agrees with that predicted by eq.~(\ref{asymptotic_IB}), 
 for all values of $F$. 

We have thus confirmed eqs.~(\ref{sumReXiIB}) 
and (\ref{asymptotic_IB}), which have been derived 
from the general results, eqs.~(\ref{sr:ReXi}) and (\ref{asymptotic}),
respectively.
Conversely, as discussed above, the agreement of our numerical results
with the general results indicates the following:
(i) this model is physically reasonable, 
and (ii) our MD simulation well describes NESSs and their responses.
That is, validity of the numerical results have been 
checked against the general results 
{\em even for large $|F|$}.
In contrast, a typical conventional method 
is to check results against the FDR (see ref.~\citen{YIS} and
references cited therein),
which holds {\em only for small $|F|$}.

{\em Summary -- }
We have derived general properties of 
linear response functions $\Phi_F$ (and their Fourier transform 
$\Xi_F$) of NESSs, which are driven by a strong field $F$.
For completeness, we have presented all the basic properties
(including rather obvious ones, such as the dispersion relations)
which correspond to those of response functions $\Phi_{\rm eq}$ 
($\Xi_{\rm eq}$) of
equilibrium states\cite{KTH,Zubarev}.
Specifically, 
we have derived the response-correlation relation, eq.~(\ref{eq:RCR}), 
which however cannot be recast into the form of 
the Kubo formula when $F \neq 0$.
As a result, the FDR and reciprocal relations are violated in NESSs, 
although the latter holds for the integrated values, 
eq.~(\ref{reciprocal}).
In contrast, 
the dispersion relations, eqs.~(\ref{eq:DR1}) and (\ref{eq:DR2}), 
and the moment sum rules hold even when $F \neq 0$
because they come from the phenomenological equations, 
eqs.~(\ref{eq:linear_response}) and (\ref{eq:causality}).
Furthermore, 
the sum rules and asymptotic behaviors,
eqs.~(\ref{sr:ReXi})-(\ref{asymptotic}),
hold even when $F \neq 0$ if 
the expectation values in an equilibrium state,
${\rm Tr} \left( \hat \rho_{\rm eq} \ \cdot \ \right)$,
are replaced with 
those in a NESS, 
${\rm Tr} \left( \hat \rho_F \ \cdot \ \right)$.
We have illustrated some of these results by an MD simulation
of an electrical conductor.

These results are quite general, which apply to diverse physical systems 
because no assumption has been made except that the NESSs are stable.
Further generalization 
to the case of time-dependent 
and/or spatially-varying $F$ (and/or $f$) 
is straightforward.

This work was supported by KAKENHI No.~19540415 
and the Grant-in-Aid for the GCOE Program 
``Weaving Science Web beyond Particle-Matter Hierarchy''.


\begin{thebibliography}{99}
\bibitem{range_of_tau}
Although the upper limit of $\tau$ is $t_{\rm out} - t_{\rm in}$, 
one can make it arbitrarily large by 
increasing the size of the source of $F$.
\end{thebibliography}
\end{document}